\documentclass[twocolumn]{jpsj2} 
\def\runauthor{Kenji {\sc Goto} {\it et al.}}

\title{Pressure-Induced Magnetic Quantum Phase Transition in Gapped Spin System KCuCl$_3$}

\author{Kenji {\sc Goto}\thanks{E-mail: goto@lee.phys.titech.ac.jp.}, Masashi {\sc Fujisawa}, Hidekazu {\sc Tanaka}$^{1}$, Yoshiya {\sc Uwatoko}$^{2}$, Akira {\sc Oosawa}$^{3}$,\\ Toyotaka \textsc{Osakabe}$^{4}$ and Kazuhisa {\sc Kakurai}$^{4}$}

\inst{Department of Physics, Tokyo Institute of Technology, Oh-okayama, Meguro-ku, Tokyo 152-8551 \\
$^1$Research Center for Low Temperature Physics, Tokyo Institute of Technology, Oh-okayama, Meguro-ku, Tokyo 152-8551 \\
$^{2}$Institute for Solid State Physics, The University of Tokyo, 
Kashiwanoha, Kashiwa, Chiba 277-8581 \\
$^{3}$Department of Physics, Sophia University, Kioi-cho, Chiyoda-ku, Tokyo 102-8554 \\
$^{4}$Quantum Beam Science Directorate, Japan Atomic Energy Agency, Tokai, Ibaraki 319-1195}

\abst{Magnetization and neutron elastic scattering measurements under a hydrostatic pressure were performed on KCuCl$_3$, which is a three-dimensionally coupled spin dimer system with a gapped ground state. It was found that an intradimer interaction decreases with increasing pressure, while the sum of interdimer interactions increases. This leads to the shrinkage of spin gap. A quantum phase transition from a gapped state to an antiferromagnetic state occurs at $P_{\rm c} \approx 8.2$ kbar. For $P > P_{\rm c}$, magnetic Bragg reflections were observed at reciprocal lattice points equivalent to those for the lowest magnetic excitation at zero pressure. This confirms that the spin gap decreases and closes under applied pressure.}

\kword{KCuCl$_3$, spin gap, pressure-induced magnetic ordering, quantum phase transition, quantum critical point, magnetization, neutron scattering}

\begin{document}
\maketitle

\section{Introduction} 
The application of magnetic field to a gapped spin singlet state can induce a quantum phase transition (QPT) to an ordered state \cite{Tachiki}. This field-induced magnetic QPT has been extensively studied in many gapped spin systems \cite{Diederix,Chaboussant,Honda,Manaka,Oosawa1,Tanaka1,Oosawa2,Rueegg,Jaime}. The results obtained can be comprehensively understood in terms of the Bose-Einstein condensation of spin triplets \cite{Giamarchi,Nikuni,Wessel,Rice,Matsumoto1,Sherman,Sirker,Nohadani1,Kawashima,Misguich}. Recently, it has been observed that the spin gap in TlCuCl$_3$ collapses under a hydrostatic pressure, so that this system can switch from a gapped singlet state to an antiferromagnetic state \cite{Goto,Oosawa3,Oosawa4,Rueegg2}. The critical pressure $P_{\rm c}$ was obtained to be $P_{\rm c}=0.42$ kbar from magnetization measurements \cite{Goto}, while $P_{\rm c}=1.07$ kbar from neutron scattering experiments \cite{Rueegg2}. The discovery of the pressure-induced QPT in TlCuCl$_3$ encouraged theoretical work, and novel phenomena dominated by quantum fluctuation were predicted for the pressure-induced QPT, e.g., the presence of the amplitude mode for magnetic excitations above $P_{\rm c}$ that has been disregarded in conventional antiferromagnets \cite{Matsumoto2} and the logarithmic correction for the gap near $P_{\rm c}$ \cite{Nohadani2}. The difference between the field- and pressure-induced QPTs is that the former arises from the softening of one of the three triplet excitation modes, while the latter arises from the simultaneous softening of the three triplet modes.

\begin{figure}[htbp]
  \begin{center}
    \includegraphics[keepaspectratio=true,width=80mm]{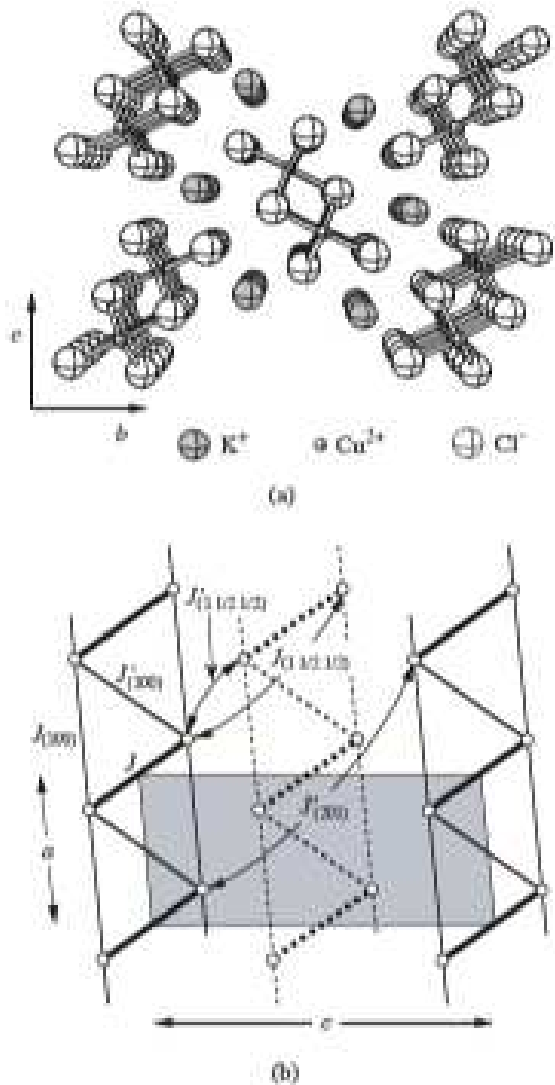}
  \end{center}
  \caption{(a) Crystal structure viewed along the $a$-axis and (b) exchange interactions in KCuCl$_3$. The double chain located at the corner and the center of the chemical unit cell in the $bc$-plane are represented by the solid and dashed lines, respectively. The shaded area shows the chemical unit cell in the $ac$-plane. The intradimer interaction is denoted as $J$, and the interdimer interaction between spins in dimers separated by the lattice vector $l{\mib a} + m{\mib b} + n {\mib c}$ is denoted as $J_{lmn}$ for pairs of spins at equivalent positions and as $J_{lmn}'$ for spins at inequivalent positions.}
  \label{fig:1}
\end{figure}
This study is concerned with a pressure-induced QPT in KCuCl$_3$, which has the same crystal structure as TlCuCl$_3$. KCuCl$_3$ has a monoclinic structure (space group $P2_1/c$) \cite{Willett}. The crystal structure is composed of planar dimers of Cu$_2$Cl$_6$. The dimers are stacked to form infinite double chains parallel to the crystallographic $a$-axis. These double chains are located at the corners and center of the unit cell in the $bc$-plane, as shown in Fig. 1(a). The magnetic ground state of KCuCl$_3$ is a spin singlet with the excitation gap $\Delta /k_{\rm B} = 30.5$ K \cite{Tanaka2,Shiramura,Oosawa2}. The lowest magnetic excitation occurs at ${\mib Q}=(0, 0, 1)$ and its equivalent reciprocal points \cite{Kato1,Cavadini1,Mueller,Cavadini2,Kato2}. The origin of the gap is the strong antiferromagnetic exchange interaction $J/k_{\rm B}\simeq 50$ K on the planar dimer Cu$_2$Cl$_6$, where the exchange interaction is defined as ${\cal H}_{\rm ex} = J_{ij}{\mib S}_i\cdot {\mib S}_j$. As shown in Fig. 1(b), neighboring dimers couple magnetically along the $a$-axis and in the $(1, 0, {\bar 2})$ plane, in which the hole orbitals of Cu$^{2+}$ spread. The gap in KCuCl$_3$ is much larger than $\Delta /k_{\rm B} = 7.5$ K in TlCuCl$_3$ \cite{Oosawa1,Shiramura}. This is because individual interdimer exchange interactions in KCuCl$_3$ are two or three times as small as those in TlCuCl$_3$ \cite{Cavadini3,Oosawa5}. In Table I, we list the individual exchange interactions in KCuCl$_3$ determined by M\"{u}ller and Mikeska \cite{Mueller}. They applied a cluster series expansion to analyze the dispersion relations observed in KCuCl$_3$ \cite{Cavadini1}.

\begin{table}[tb]
\caption{Exchange interactions in KCuCl$_3$ in K units \cite{Mueller}.
\label{table1}}
\begin{center}
\begin{tabular}{cccccc} \hline
$J$ & $J_{(100)}$ & $J'_{(100)}$ & $J_{\left(1,\frac{1}{2},\frac{1}{2}\right)}$ & $J'_{\left(1,\frac{1}{2},\frac{1}{2}\right)}$ & $J'_{(2,0,1)}$ \\ \hline
49.3 & $-0.24$ & 4.93 & 9.86 & 1.97 & 9.27 \\ \hline
\end{tabular}
\end{center}
\end{table}

Within the framework of the dimer mean-field approximation \cite{Oosawa2}, the gap is expressed as 
$$
\Delta=\sqrt{J(J - 2|{\tilde J}|)} , \eqno (1)
$$ 
where ${\tilde J}$ is expressed by a certain linear combination of interdimer interactions, and for the present system, 
$$
{\tilde J} = J_{(100)} - \frac{1}{2}J'_{(100)} - J_{\left(1,\frac{1}{2},\frac{1}{2}\right)} + J'_{\left(1,\frac{1}{2},\frac{1}{2}\right)} - \frac{1}{2}J'_{(2,0,1)},  \eqno (2)
$$
where individual interdimer interactions are shown in Fig. 1(b). The gap shrinks either when the intradimer interaction $J$ is reduced or when the interdimer interaction is enhanced by an applied pressure. In the case of TlCuCl$_3$, we were not able to distinguish which effect is dominant because of the small critical pressure $P_{\rm c}$. On the other hand, the gap in KCuCl$_3$ is approximately four times as large as the gap in TlCuCl$_3$, and thus, $P_{\rm c}$ for KCuCl$_3$ should be much larger than that for TlCuCl$_3$. Therefore, we can obtain the pressure dependence of both intradimer and interdimer interactions before reaching the critical pressure. With this reasoning, we performed magnetization measurements and neutron elastic scattering experiments on KCuCl$_3$ under a hydrostatic pressure.

\section{Experimental}
Single KCuCl$_3$ crystals were prepared by the Bridgman method. The temperature of the center of the furnace was set at 500$^{\circ}$C. The details of sample preparation are almost the same as those for TlCuCl$_3$ \cite{Oosawa1}. KCuCl$_3$ crystals are easily cleaved along the $(1, 0, {\bar 2})$ plane. The second cleavage plane is $(0, 1, 0)$. These cleavage planes are perpendicular to each other and parallel to the $[2,0,1]$ direction.

Magnetizations were measured at temperatures down to 1.8 K under magnetic fields up to 7 T using a SQUID magnetometer (Quantum Design MPMS XL). Pressures up to 10 kbar were applied using a cylindrical high-pressure cramp cell designed for use with the SQUID magnetometer \cite{Uwatoko}. A sample with a size of $\sim 2.5\times 2.5\times 5$ mm$^3$ was set in the cell with its $[2, 0, 1]$ direction parallel to the cylindrical axis. A magnetic field was applied along the $[2, 0, 1]$ direction. The $g$-factor for this field direction is $g=2.04$, which we obtained by ESR measurement. Daphne oil 7373 and liquid paraffin were used as pressure-transmitting fluids. The pressure was calibrated with the superconducting transition temperature $T_{\rm c}$ of tin placed in the pressure cell. The diamagnetism of tin was measured at $H=10$ and 50 Oe to determine $T_{\rm c}$ after removing the residual magnetic flux trapped in the superconducting magnet. The accuracy of the pressure is 0.1 kbar for the absolute value.

For neutron scattering experiments, a McWhan-type high-pressure cell (HPCNS-MCW, Oval Co., Ltd.) \cite{McWhan} was used. The sample with a volume of $\sim 0.2$ cm$^3$ was set in the high-pressure cell. A mixture of Fluorinert FC70 and FC77 was used as pressure-transmitting medium. The hydrostatic pressures $P=11$, 14 and 21 kbar were applied at helium temperatures. The pressure was determined from the pressure dependence of the lattice constants of a NaCl crystal placed in the sample space. Neutron elastic scattering measurements were performed using a TAS-1 spectrometer installed at JRR-3M in JAEA, Tokai. The incident neutron energy was fixed at $E_i = 14.7$ meV. Because the size of the sample has to be small due to the small sample space, collimations were set as open$-80'-80'-80'$ to gain intensity. Sapphire and pyrolytic graphite filters were placed to suppress the background by high energy neutrons, and higher order contaminations, respectively. The sample was mounted in the cryostat with its $a^*$- and $c^*$-axes in the scattering plane. Table II shows the lattice parameters $a$, $c$ and $\beta$ at $P=11$, 14 and 21 kbar. These three lattice parameters monotonically decreases with increasing pressure, although a nonlinear pressure dependence is observed for $a$ and $\beta$.

\begin{table}[tb]
\caption{Pressure dependences of lattice parameters $a$, $c$ and $\beta$.\label{table2}}
\begin{center}
\begin{tabular}{clccccc} \hline
& P [kbar] & 0  & 11  & 14  & 21 &  \\ \hline
& $a$ [{\AA}] & 4.029 & 3.886 & 3.863 & 3.807 &\\
& $c$ [{\AA}] & 8.736 & 8.492 & 8.434 & 8.283 &\\
& $\beta$ [deg] & 97.33 & 95.73 & 95.50 & 95.08 &\\ \hline
\end{tabular}
\end{center}
\end{table}

\section{Results and Discussion}
\subsection{Magnetization}
\begin{figure}[htbp]
  \begin{center}
    \includegraphics[keepaspectratio=true,width=85mm]{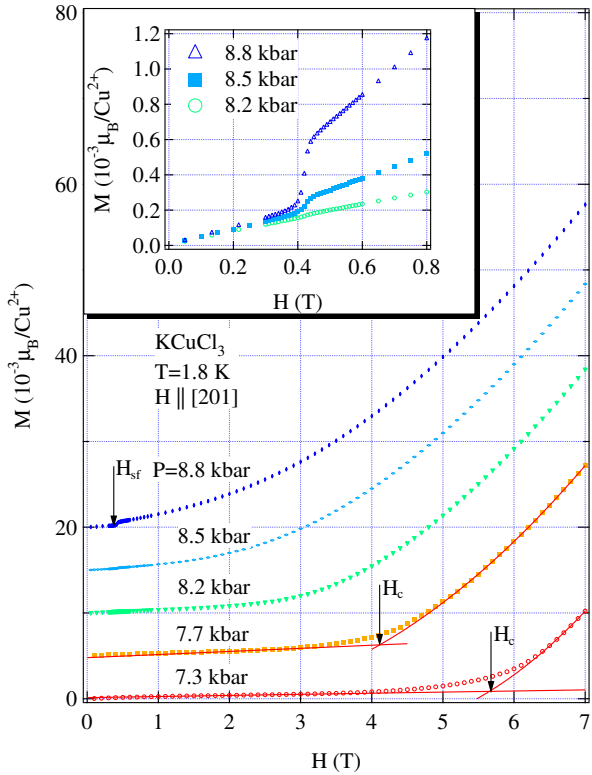}
  \end{center}
  \caption{Magnetization curves for KCuCl$_3$ measured at $T$ = 1.8 K under various pressures. The magnetic field is applied parallel to the $[2, 0, 1]$ direction. For clarity, magnetizations are shifted upward consecutively by $5\times 10^{-3}\ \mu_{\rm B}$ with increasing pressure. Arrows indicate the critical fields $H_{\rm c}$. The inset shows the magnetization curve for $H \leq 0.8$ T measured at $P=8.2$, 8.5 and 8.8 kbar.}
  \label{fig:3}
\end{figure}

At an ambient pressure, magnetization is almost zero at $T=1.8$ K for $H\leq 7$ T, because the critical field $H_{\rm c}={\it \Delta}/g\mu_{\rm B}$ corresponding to the gap is $(g/2)H_{\rm c}\approx 23$ T in KCuCl$_3$ \cite{Oosawa2,Shiramura}. However, for $P>7$ kbar, a rapid increase in magnetization is observed below 7 T due to the shrinkage of spin gap. Figure \ref{fig:3} shows the magnetization curves measured at $T$=1.8 K under various pressures for $P \geq 7.3$ kbar. Arrows indicate the critical fields $H_{\rm c}$, at which the excitation gap closes. To evaluate the critical fields $H_{\rm c}$, we used two fitting functions for magnetizations, the linear function ($K_0+K_1H$) for $H < H_{\rm c}$ and the quadratic function ($K_0'+K_1'H+K_2'H^2$) for $H > H_{\rm c}$. These fitting functions are represented by solid lines in Fig. \ref{fig:3}. The critical fields indicated by arrows in Fig. \ref{fig:3} were evaluated from a field at which the two fitting functions cross. The magnetization of the ground state is proportional to $H-H_{\rm c}$ just above $H_{\rm c}$, and the slope of the magnetization curve decreases with decreasing $H_{\rm c}$ \cite{Matsumoto2}. For this reason and the finite temperature effect together with the small inhomogeneity of the pressure, the bend anomaly at $H_{\rm c}$ observed at 1.8 K becomes smeared with increasing pressure. Therefore, there is a certain amount of error in the determination of $H_{\rm c}$. 

It is evident that the gap shrinks with hydrostatic pressure and vanishes at around 8.2 kbar. As shown in the inset of Fig. \ref{fig:3}, a spin-flop transition is clearly observed for $P=8.8$ kbar at $H_{\rm sf} = 0.42$ T. This is typical of the antiferromagnetic ordering with the easy-axis close to the applied field direction. A small spin-flop transition is also observed for $P=8.5$ kbar, while not for $P=8.2$ kbar. From the present results, we observe that KCuCl$_3$ undergoes a pressure-induced QPT from a gapped ground state to an antiferromagnetic state at $P_{\rm c}\approx 8.2$ kbar. The critical pressure in KCuCl$_3$ is about ten times as large as $P_{\rm c}=0.42\sim 1.07$ kbar in TlCuCl$_3$ \cite{Goto,Rueegg2}. 

\begin{figure}[htbp]
  \begin{center}
    \includegraphics[keepaspectratio=true,width=80mm]{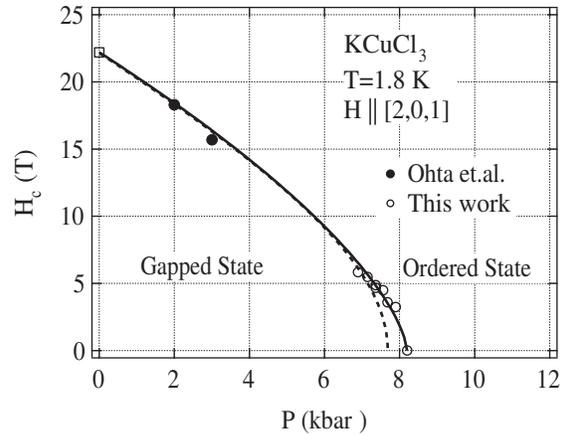}
  \end{center}
  \caption{Pressure dependence of critical field $H_{\rm c}$. The solid line serves as a visual guide. The dashed line shows the calculation results obtained using eq. (1) with $J/k_{\rm B}=48.9-1.81P$ K and $|{\tilde J}|=15.0-0.327P$ K.}
  \label{fig:4}
\end{figure}

The pressure dependence of $H_{\rm c}$ is plotted in Fig. \ref{fig:4}, where we added $H_{\rm c}=22.2$ T obtained by the previous high-field magnetization measurement at an ambient pressure \cite{Oosawa2}. The singlet-triplet excitation corresponding to the spin gap in KCuCl$_3$ can be observed by ESR due to the presence of the weak antisymmetric exchange interaction of the Dzyaloshinsky-Moriya type that has matrix elements between the singlet ground state and the lowest excited triplet \cite{Tanaka3,Budhy}. Recently, Ohta and his coworkers \cite{Ohta} have investigated the pressure dependence of the gap in KCuCl$_3$ up to 3.2 kbar by means of high-field and high-frequency ESR. Their result is in accordance with our result, as shown in Fig. \ref{fig:4}.

\begin{figure}[htbp]
  \begin{center}
    \includegraphics[keepaspectratio=true,width=80mm]{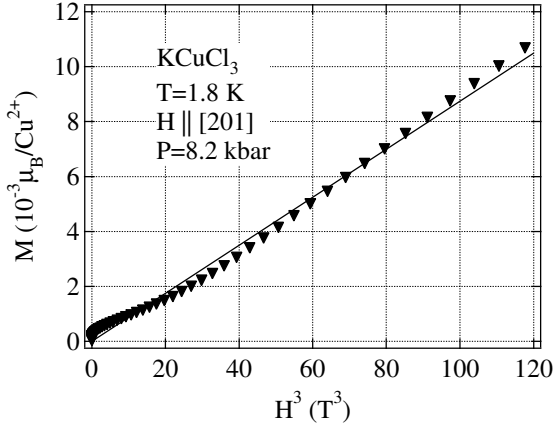}
  \end{center}
  \caption{Magnetization vs $H^3$ at $P=8.2$ kbar measured at $T=1.8$ K. The solid line denotes the fit obtained using eq. (3) with $J/k_{\rm B}= 31.0$ K and $g=2.04$.}
  \label{fig:5}
\end{figure}

As shown in Fig. \ref{fig:3}, the magnetization curve at $P=8.2$ kbar that approximates the critical pressure is largely rounded in the low-field region. At $P=P_{\rm c}$, two triplet components, $\left|1,1\right>$ and $\left|1,-1\right>$, equally contribute to the ground state at zero field. Here, $|S,S^z\rangle$ denotes the spin state of a dimer with the total spin $S$ and the $z$ component $S^z$. In the finite external field $H$, the amplitude of $\left|1,1\right>$ is enhanced, while that of $\left|1,-1\right>$ is suppressed. For this reason, the low-field magnetization at $P=P_{\rm c}$ is proportional to $H^3$. The magnetization per spin $m$ is given by \cite{Matsumoto2}
$$
m\simeq (g/2)\mu_{\rm B}\left(g\mu_{\rm B}H/J\right)^3. \eqno (3)
$$
In Fig. \ref{fig:5}, we plot the magnetization at $P=8.2$ as a function of $H^3$. We can see that the magnetization is approximately proportional to $H^3$, as theoretically predicted. The intradimer exchange interaction evaluated by applying eq. (3) to the experimental data shown in Fig. \ref{fig:5} is $J/k_{\rm B}= 31.0$ K. This $J$ is somewhat smaller than $J/k_{\rm B}= 34.4$ K obtained from the analysis of the magnetic susceptibility data as described below.

\begin{figure}[htbp]
  \begin{center}
    \includegraphics[keepaspectratio=true,width=80mm]{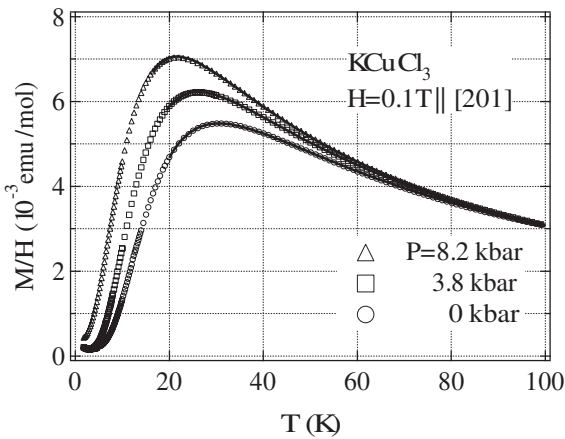}
  \end{center}
  \caption{Temperature dependence of magnetic susceptibilities $\chi$ in KCuCl$_3$ measured at $P=0$, 3.8 and 8.2 kbar for $H\parallel [2, 0, 1]$. Solid lines denote the fits obtained using eq. (4).}
  \label{fig:6}
\end{figure}

We measured the magnetic susceptibilities of KCuCl$_3$ at various pressures to investigate the systematic change in exchange interaction. Figure \ref{fig:6} shows the temperature dependence of the susceptibilities ($\chi=M/H$) measured at the pressures of 0, 3.8 and 8.2 kbar. A magnetic field of 0.1 T was applied. With decreasing temperature, the susceptibilities exhibit broad maxima and rapidly decrease toward zero. This behavior is characteristic of the gapped spin system. We can see that with increasing pressure, the temperature $T_{\rm{max}}$ giving the susceptibility maximum $\chi_{\rm{max}}$ decreases and that the value of $\chi_{\rm{max}}$ increases. 

\begin{figure}[htbp]
  \begin{center}
    \includegraphics[keepaspectratio=true,width=80mm]{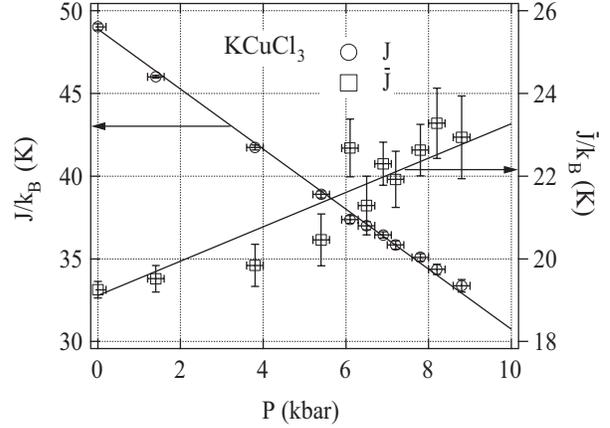}
  \end{center}
  \caption{Pressure dependence of intradimer interaction $J$ and sum of interdimer interactions $\bar J$.}
  \label{fig:7}
\end{figure}

When interdimer exchange interactions are treated as mean-fields, the magnetic susceptibility $\chi$ is expressed by 
$$
\chi=\frac{2g^2\mu_{\rm B}^2\beta N}{3+e^{\beta J}+2\beta {\bar J}}, \eqno (4)
$$ 
where $\beta=1/(k_{\rm B}T)$, $N$ is the number of dimers and 2${\bar J}$ is the sum of interdimer interactions acting on one spin in a dimer, which is given for the present system by
$$
{\bar J} = J_{(100)} + \frac{1}{2}J'_{(100)} + J_{\left(1,\frac{1}{2},\frac{1}{2}\right)} + J'_{\left(1,\frac{1}{2},\frac{1}{2}\right)} + \frac{1}{2}J'_{(2,0,1)}.  \eqno (5)
$$
In the present mean field approximation, $T_{\rm{max}}$ is given by $J$ only as $1.60T_{\rm{max}}=J/k_{\rm B}$. Solid lines in Fig. \ref{fig:6} denote the fits obtained using eq. (4) for $T > (2/3)T_{\rm{max}}$ with $g=2.04$. The $J$ and ${\bar J}$ obtained by fitting are plotted in Fig. \ref{fig:7}.  
From Fig. \ref{fig:7}, it is evident that with increasing pressure, the intradimer interaction $J$ decreases, while the sum of interdimer interactions ${\bar J}$ increases. $J$ exhibits a linear pressure dependence and is expressed as
$$
J/k_{\rm B}=(48.9-1.81P)\ {\rm K}, \eqno (6)
$$
where the unit of pressure is kbar. Assuming that ${\bar J}$ also exhibits a linear pressure dependence, we obtained  
$$
{\bar J}/k_{\rm B}=(19.1+0.416P)\ {\rm K}. \eqno (7)
$$ 
Solid lines in Fig. \ref{fig:7} denote these pressure dependences. $J/k_{\rm B}=48.9$ K obtained at an ambient pressure is consistent with $J/k_{\rm B} = 49.3$ K obtained from the analysis of the dispersion relation obtained by neutron inelastic scattering experiments \cite{Cavadini1,Mueller,Cavadini2,Kato2}. 

From the present susceptibility measurements, it was found that the intradimer interaction decreases with increasing pressure, while the interdimer interaction increases. From eq. (1), we see that these pressure dependences act cooperatively to suppress the spin gap. This is why the spin gap in KCuCl$_3$ shrinks with increasing pressure. The same must be true of TlCuCl$_3$. It is considered that with increasing pressure, the bond angle of the intradimer exchange path $\rm{Cu^{2+}-Cl^{-}-Cu^{2+}}$ becomes closer to 90$^{\circ}$, so that the antiferromagnetic intradimer interaction is suppressed. The enhancement in interdimer interaction should be attributed to the contraction of the interdimer distance.

On the assumption that ${\tilde J}$ in eq. (2) has the same pressure dependence as ${\bar J}$ given by eq. (7), i.e., $|{\tilde J}|=\left(15.0 +0.327P\right)$, we calculated the critical field $H_{\rm c}$ with $J$ given by eq. (6). The dashed line in Fig. \ref{fig:4} denotes the calculated critical field. The calculated critical pressure $P_{\rm c}=7.7$ kbar is slightly smaller than the experimental $P_{\rm c}\approx 8.2$ kbar. This disagreement should be mainly due to the error in the estimation of ${\bar J}$.

\begin{figure}[htbp]
  \begin{center}
    \includegraphics[keepaspectratio=true,width=80mm]{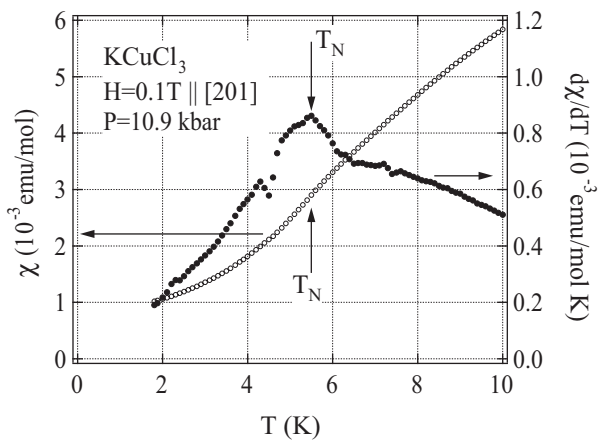}
  \end{center}
  \caption{Low-temperature magnetization $M/H$ and $dM/dT$ in KCuCl$_3$ measured at $P=10.9$ kbar for $H\parallel [2, 0, 1]$. The arrow denotes the ordering temperature $T_{\rm N}$.}
  \label{fig:8}
\end{figure}

Figure  \ref{fig:8} shows the low-temperature susceptibility $\chi$ and its temperature derivative ${\rm d}{\chi}/{\rm d}T$ measured at $P=10.9$ kbar. For $P > 10$ kbar, magnetic susceptibility exhibits a small inflectional anomaly due to magnetic ordering, as shown in Fig. \ref{fig:8}. The small anomaly at 4.4 K is due to an instrumental problem and is not intrinsic to the sample. We assign the temperature giving the inflection point of magnetization to the ordering temperature $T_{\rm N}$. The pressure dependence of $T_{\rm N}$ obtained from the magnetization measurement and neutron elastic scattering experiment is discussed in the next subsection.

\subsection{Neutron elastic scattering}
\begin{figure}[htbp]
  \begin{center}
    \includegraphics[keepaspectratio=true,width=80mm]{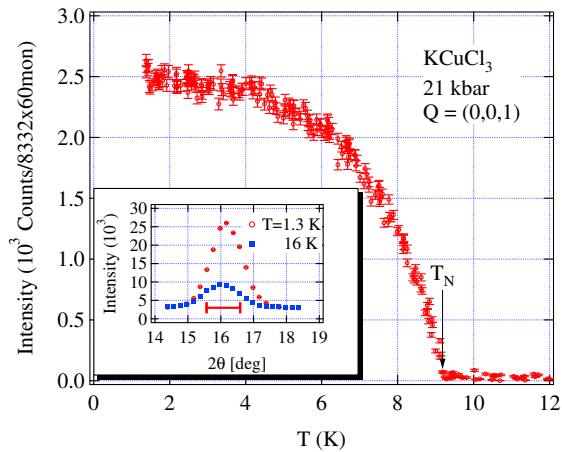}
  \end{center}
  \caption{Temperature dependence of magnetic Bragg peak intensity for ${\mib Q}=(0, 0, 1)$ reflection measured at $P=21$ kbar in KCuCl$_3$. The inset shows ${\theta}-2{\theta}$ scans for the ${\mib Q}=(0, 0, 1)$ reflection measured at $T=1.3$ and 16 K for $P=21$ kbar. The horizontal bar denotes the calculated instrumental resolution width.}
  \label{fig:9}
\end{figure}

To confirm the pressure-induced magnetic ordering in KCuCl$_3$, we performed neutron elastic scattering under the pressures of $P=11$, 14 and 21 kbar. With decreasing temperature, magnetic Bragg reflections with a resolution-limited width were observed at ${\mib Q}=(h, 0, l)$ with the integer $h$ and odd $l$, which are equivalent to those for the lowest magnetic excitation at an ambient pressure. The inset in Fig. \ref{fig:9} shows the ${\theta}-2{\theta}$ scans for the ${\mib Q} = (0, 0, 1)$ reflection measured at $T=1.3$ and 16 K for $P=21$ kbar. The ordering temperature at this pressure is $T_{\rm N}=9.2$ K. The weak peak observed at $T=16$ K is a nuclear peak, which is forbidden for the space group $P2_1/c$. This weak nuclear peak may be ascribed to the higher order contamination of the strong $(0, 0, 2)$ nuclear reflection or the local distortion of a lattice due to the applied pressure. Hence, we subtracted the nuclear contribution from the scattering data to obtain a genuine magnetic contribution. Figure \ref{fig:9} shows the temperature dependence of the magnetic peak intensity at $\mib Q=(0, 0, 1)$ measured at $P=21$ kbar. A phase transition is clearly observed at $T_{\rm N}=9.2$ K. These results indicate that the spin gap closes at $P_{\rm c}\approx 8.2$ kbar and that for $P > P_{\rm c}$, the ground state is gapless. 

\begin{figure}[htbp]
  \begin{center}
    \includegraphics[keepaspectratio=true,width=80mm]{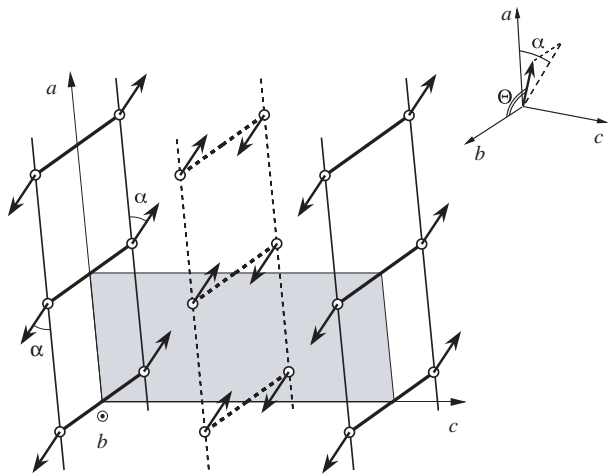}
  \end{center}
  \caption{Spin structure of pressure-induced ordered phase in KCuCl$_3$ projected onto the $ac$-plane. The inset shows the definitions of the angles $\alpha$ and ${\varTheta}$ representing the spin direction.}
  \label{fig:10}
\end{figure}

We measured the integrated intensities of nine Bragg reflections at $T=1.3$ K to determine the spin structure of the pressure-induced ordered phase.
Table \ref{table3} summarizes the integrated intensities of magnetic Bragg reflections normalized to the $(0, 0, 1)$ reflection. These reciprocal lattice points are equivalent to those for the pressure-induced magnetic ordering in TlCuCl$_3$ \cite{Oosawa3,Oosawa4}. Thus, we fit the observed magnetic peak intensities with the magnetic structure similar to that of the pressure-induced ordered phase in TlCuCl$_3$. The magnetic structure is characterized by the antiferromagnetic ordering within the dimer, ferromagnetic ordering along the double chain and antiferromagnetic spin arrangement between the nearest neighbor chains, as shown in Fig. \ref{fig:10}. We introduce two angles, $\alpha$ and $\varTheta$, to represent the spin direction, as shown in the inset in Fig. \ref{fig:10}. The spin direction in the $ac$-plane is expressed by the angle $\alpha$ to the $a$-axis. The out-of-plane spin direction is expressed by the angle $\varTheta$ from the $b$-axis.
To refine the magnetic structure, we used the atomic coordinates of KCuCl$_3$ at an ambient pressure \cite{Willett} and the nuclear scattering lengths $b_{\rm K}=3.71$, $b_{\rm Cu}=7.72$ and $b_{\rm Cl}=9.58$ with 10$^{-13}$ cm unit \cite{Sears}. The magnetic form factors of Cu$^{2+}$ were taken from ref. \citen{Brown}. The extinction effect was evaluated by comparing the observed and calculated intensities for various nuclear Bragg reflections. The calculated intensities with these two angles as fitting parameters are shown in Table \ref{table3}. The calculated intensities are in agreement with the observed ones. The angle $\varTheta$ is close to $90^{\circ}$ for $P=11$ and 14 kbar, while for $P=21$ kbar, $\varTheta$ is clearly much smaller than $90^{\circ}$. This indicates that the ordered spins lie almost in the $ac$-plane up to $P=14$ kbar and that they incline toward the $b$-axis at $P=21$ kbar. This ordered spin inclination toward the $b$-axis may be indicative of the spin reorientation on lowering temperature as has been observed in the pressure induce-ordered phase of TlCuCl$_3$ \cite{Oosawa3,Oosawa4}.

\begin{fulltable}[tb]
\caption{Observed and calculated magnetic Bragg peak intensities measured at $T=1.8$ K for $P=11$, 14 and 21 kbar in KCuCl$_3$. The intensities are normalized to the (0, 0, 1)$_{\rm M}$ reflection. $R$ is the reliability factor given by $R=\sum_{h,k,l}|I_{\rm cal}-I_{\rm obs}|/\sum_{h,k,l}I_{\rm obs}$.\label{table3}}
\begin{center}
\begin{tabular}{cccccccccc} \hline
 & & \multicolumn{2}{c}{$P=11$ kbar} &    & \multicolumn{2}{c}{$P=14$ kbar}&    & \multicolumn{2}{c}{$P=21$ kbar}\\ \hline
$(h, k, l)$ & \hspace{0.2cm} & $I_{\rm obs}$ & $I_{\rm cal}$  & \hspace{0.2cm} & $I_{\rm obs}$ & $I_{\rm cal}$  & \hspace{0.2cm} &$I_{\rm obs}$ & $I_{\rm cal}$ \\ \hline
(0, 0, 1)$_{\rm M}$ & \hspace{0.2cm} & 1 $\pm$ 0.018 & 1 & \hspace{0.2cm} & 1 $\pm$ 0.493 & 1 & \hspace{0.2cm} & 1 $\pm$ 0.018&1\\
(0, 0, 3)$_{\rm M}$ & \hspace{0.2cm} & 0 $\pm$ 0.008 & 0.009 & \hspace{0.2cm} & 0 $\pm$ 0.002 & 0.009 & \hspace{0.2cm} & 0 $\pm$ 0.002&0.008\\
(0, 0, 5)$_{\rm M}$ & \hspace{0.2cm} & 0.066 $\pm$ 0.012 & 0.089 & \hspace{0.2cm} & 0.089 $\pm$ 0.007 & 0.088 & \hspace{0.2cm} & 0.086 $\pm$ 0.004&0.087\\
(1, 0, 1)$_{\rm M}$ & \hspace{0.2cm} & 0.095 $\pm$ 0.015 & 0.093 & \hspace{0.2cm} & 0.094 $\pm$ 0.011 & 0.106 & \hspace{0.2cm} & 0.120 $\pm$ 0.007& 0.187\\
(1, 0, $-1$)$_{\rm M}$ & \hspace{0.2cm} & 0.030 $\pm$ 0.014 & 0.019 & \hspace{0.2cm} & 0.009 $\pm$ 0.007 & 0.008 & \hspace{0.2cm} & 0.152 $\pm$ 0.008& 0.152\\
(1, 0, 3)$_{\rm M}$ & \hspace{0.2cm} & 0.310 $\pm$ 0.024 & 0.260 & \hspace{0.2cm} & 0.214 $\pm$ 0.017 & 0.285 & \hspace{0.2cm} & 0.273 $\pm$ 0.008 & 0.301\\
(1, 0, $-3$)$_{\rm M}$ & \hspace{0.2cm} & 0.055 $\pm$ 0.016 & 0.056 & \hspace{0.2cm} & 0.042 $\pm$ 0.004 & 0.040 & \hspace{0.2cm} & 0.120 $\pm$ 0.003 & 0.129\\
(2, 0, 1)$_{\rm M}$ & \hspace{0.2cm} & 0.068 $\pm$ 0.020 & 0.057 & \hspace{0.2cm} & 0.076 $\pm$ 0.005 & 0.064 & \hspace{0.2cm} & 0.193 $\pm$ 0.004&0.154\\
(2, 0, $-1$)$_{\rm M}$ & \hspace{0.2cm} & 0.018 $\pm$ 0.010 & 0.012 & \hspace{0.2cm} & 0.006 $\pm$ 0.002 & 0.009 & \hspace{0.2cm} & 0.108 $\pm$ 0.003 &0.081\\ \hline
$\alpha$ (deg)& \hspace{0.2cm} & & 29.9 $\pm$ 2.3 & \hspace{0.2cm} & & 33.2 $\pm$ 1.1 & \hspace{0.2cm} & & 61.2 $\pm$ 1.1\\
$\Theta$  (deg)& \hspace{0.2cm} & & 71.7 $\pm$ 4.6 & \hspace{0.2cm} & & 79.5 $\pm$ 2.7 & \hspace{0.2cm} & &45.7 $\pm$ 0.8\\ \hline
$R$ & \hspace{0.2cm} & & 0.08 & \hspace{0.2cm} & & 0.08 & \hspace{0.2cm} & &0.12\\ \hline
\end{tabular}
\end{center}
\end{fulltable}

\begin{figure}[htbp]
  \begin{center}
    \includegraphics[keepaspectratio=true,width=80mm]{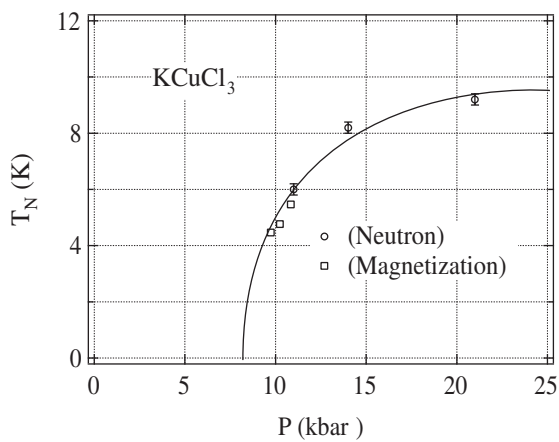}
  \end{center}
  \caption{Pressure dependence of ordering temperature $T_{\rm N}$. The solid line serves as a visual guide.}
  \label{fig:11}
\end{figure}

The ordered moment magnitudes evaluated at $T=1.3$ K for $P=11$, 14 and 21 kbar are $\langle m \rangle=(0.45\pm 0.02)\mu_{\rm B}$, $(0.56\pm 0.03)\mu_{\rm B}$ and $(0.91\pm 0.09)\mu_{\rm B}$, respectively. The ordered moments $\langle m\rangle$ monotonically increase with applied pressure. In Fig. \ref{fig:11}, we plotted the data of $T_{\rm N}$ obtained from the magnetization and neutron scattering measurements. The ordering temperature $T_{\rm N}$ is not proportional to $\langle m\rangle$. At a high pressure, the increase in $T_{\rm N}$ is suppressed. This behavior may be ascribed to the reduction in anisotropy energy which determines the ordered moment direction.

Finally, we evaluate the contribution of the triplet component to the ground state in the pressure induced ordered state. When the $x$-direction is considered to be parallel to the ordered moments, the basis state of the dimer $\psi$ may be expressed by 
$$
\psi=\left|0,0\right> \cos{\theta}+\frac{1}{\sqrt{2}}\left(\left|1,1\right> - \left|1,-1\right> \right)\sin{\theta},\eqno (8) 
$$
where the angle $\theta$ is introduced to satisfy the normalization condition \cite{Oosawa2}. From eq. (8), the magnitude of sublattice magnetization $\langle m \rangle$ is given by $\langle m \rangle=g\mu_{\rm B}\langle S_x \rangle=g\mu_{\rm B}\cos{\theta}\sin{\theta}$. With $\langle m \rangle=0.45$, 0.56 and 0.91$\mu_{\rm B}$ obtained at $P=11$, 14 and 21 kbar, and $g=2.07$, 2.06 and 2.05 corresponding to the directions of the ordered moments at these pressures, we obtain $\sin^2{\theta}=0.050$, 0.080 and 0.27, respectively. These values denote the contribution of the triplet component to the ground state.

\section{Conclusion}
We have presented the results of magnetization measurements and neutron elastic scattering experiments on the gapped spin system KCuCl$_3$ under a hydrostatic pressure. The gap decreases with applied pressure. The pressure dependence of the gap was obtained, as shown in Fig. \ref{fig:4}. A pressure-induced quantum phase transition from a gapped state to an antiferromagnetic state occurs at the critical pressure $P_{\rm c} \approx 8.2$ kbar. Above $P_{\rm c}$, Bragg reflections indicative of magnetic ordering were observed at ${\mib Q}=(h, 0, l)$ with the integer $h$ and odd $l$, which are equivalent to those for the lowest magnetic excitation at zero pressure. This confirms that the pressure-induced magnetic ordering observed arises from the closing of the spin gap. From the analysis of magnetic susceptibility data, it was found that an intradimer interaction decreases with increasing pressure, while sum of interdimer interactions increases. These pressure dependences of the exchange interactions give rise to reduction in spin gap. The pressure dependence of the transition temperature was obtained, as shown in Fig. \ref{fig:11}. 

\section*{Acknowledgment}
The authors would like to thank T. Ono for his technical advice on magnetization measurements.
This work was supported by a Grant-in-Aid for Scientific Research and a 21st Century COE Program at Tokyo Tech ``Nanometer-Scale Quantum Physics'' both from the Ministry of Education, Culture, Sports, Science and Technology of Japan. A. O. was supported by the Saneyoshi Scholarship Foundation and the Kurata Memorial Hitachi Science and Technology Foundation.

\end{document}